**Quantum Phase Transitions and Fractional Quantized Anomalous Hall Insulators in Rhombohedral Graphene**

Zach Hadjri[1†], Xinlei Yue[2†], Tonghang Han[1], Yuxuan Yao[1], Zhengguang Lu[3], Shenyong Ye[1], Junseok Seo[1], Jixiang Yang[1], Kenji Watanabe[4], Takashi Taniguchi[5], Liang Fu[1], Ady Stern[2]* & Long Ju[1]*

[1]Department of Physics, Massachusetts Institute of Technology, Cambridge, MA, USA.

[2]Department of Condensed Matter Physics, Weizmann Institute of Science, Rehovot, Israel

[3]Department of Physics, Florida State University, Tallahassee, FL, USA

[4]Research Center for Electronic and Optical Materials, National Institute for Materials Science, 1-1 Namiki, Tsukuba, Japan

[5]Research Center for Materials Nanoarchitectonics, National Institute for Materials Science, 1-1 Namiki, Tsukuba, Japan

*Corresponding authors. Email: longju@mit.edu, adiel.stern@weizmann.ac.il

†These authors contributed equally to this work.

**Fractional quantum anomalous Hall effect (FQAHE) has been discovered in twisted $MoTe_2$ and rhombohedral graphene/hBN moiré superlattices. Such van der Waals heterostructures feature a tuning knob of gate displacement field *D*, which is absent from the conventional fractional quantum Hall systems in two-dimensional electron gases. *D* plays a critical role in engineering FQAHE and other emergent quantum states and provides an exciting new opportunity to explore their quantum phase transitions. However, the microscopic details of such transitions and temperature-dependent transport have remained mostly elusive. Here we report systematic resistance measurements in rhombohedral pentalayer graphene/hBN moiré superlattices. We found that the displacement field-driven phase transitions between Composite Fermi liquid, Fermi liquid, Fractional Chern insulators, and insulating states are described by semi-circle relations of the longitudinal and transverse resistivities (or conductivities), largely unexplored in the fractional quantum Hall systems. This agrees with a spatially separated two-phase picture for the phase transitions and further indicates a new insulator phase—fractional quantized anomalous Hall insulator. By comparing the temperature-dependence of longitudinal resistance with the thermal activation model, we estimated the transport gap sizes in three fractional Chern insulator states. Our work shed** **light on the quantum and temperature evolutions of fractional Chern insulator states—providing necessary background for anyon-braiding and gate-defined junctions in rhombohedral graphene.**

Crystalline graphene in the rhombohedral stacking order (referred to as rhombohedral graphene or RG in the following) has developed into a new arena of intertwined correlated and topological electron physics. Its highly gate-tunable flat bands and Berry curvatures drive a plethora of emergent quantum phenomena, ranging from orbital magnetism and multiferroics to unconventional superconductivity, integer quantum anomalous Hall effect and Chern insulators[1-11]. When proximitized by the moiré superlattice from a nearly aligned hexagonal boron nitride (hBN) substrate, fractional quantum anomalous Hall effect (FQAHE)[12-15] has been observed when conduction band electrons are electrically polarized away from the moiré interface[16]. Additionally, fractional Chern insulators (FCIs)[17,18] were revealed in the same RG/hBN system when electrons are polarized towards the moiré interface at down to 0.2 T—almost the zero magnetic field that is required to show FQAHE[19]. Many questions were triggered by this rich zoo of FQAHE/FCI observations but have remained open so far[20-26]. Especially, thanks to the gate-tunable displacement field $D$—an experimental knob that is absent from conventional fractional quantum Hall systems—many other intriguing ground states have been observed in the same devices, such as valley-symmetry-breaking Fermi liquid (FL), correlated insulators (CI), extended quantum anomalous Hall (EQAH) states[27], and the likely composite Fermi liquid (CFL)[28-30]. However, their connections with FCIs by quantum phase transitions have not been fully understood. These facts kept RG/hBN system as a thriving frontier to explore many-body interaction of electrons with plenty of mysteries and opportunities.

Electron transport experiments, albeit a global measurement of resistance and more sensitive to disorder than thermodynamic and scanning probe measurements, can provide valuable insights into the understanding of many outstanding questions about FCIs. In quantum Hall systems, it has been shown that two adjacent states are connected by a semi-circle on the $\rho_{xy} - \rho_{xx}$ plane (corresponding to the macroscopic transverse and longitudinal resistivities, respectively)—a very general result that can be derived for two-phase scenarios solely based on Kirchoff-Ohm laws[31-38]. This observation suggested the two integer quantum Hall liquids occupy separated spatial regions that expand/shrink in area as the Landau level filling factor is continuously changed. Such microscopic evolution of a quantum phase transition could be hard to image by scanning probe measurements, which might not have the necessary spatial resolution or be unable to reach electron systems embedded beneath the sample surface. In addition, the temperature dependence of $R_{xx}$ could distinguish between (fractional) quantum Hall liquid and Fermi Liquid and allow for the extraction of energy gaps of the former at fractional filling factors of Landau levels and moiré flat bands, similar to what has been done in the fractional quantum Hall effect (FQHE) in GaAs/AlGaAs heterostructures[39-47], Si MOSFETs[48], and graphene[49,50].

Here we report resistance measurements in a pentalayer RG/hBN device, in which we observed FQAHE previously when electrons are polarized away from the moiré superlattice interface. With extensive efforts on improving the electronic filtering and optimization of grounding conditions of the measurement setup, we have reduced the residue resistance to hundreds of Ohms at the base temperature—about 15 times lower than that in the first observation of FQAHE in the same device[16]. This allows us to observe a total of 11 fractional states (see Extended Data Fig. 3). At fixed filling factors corresponding to FCI states, we measured the $R_{xx}$ and $R_{xy}$ as a function of $D$. We found that the quantum phase transitions between FL, FCI and EQAH states can be well-described by the semi-circles of resistivities, using a reasonable effective aspect ratio of the device. The transition between FCI and CI, on the other hand, can be well-described by the semi-circles of conductivities, suggesting a previously unidentified fractional quantized anomalous Hall insulator phase. We further compared the temperature-dependent resistance data to a thermal activation model, and estimated a largest gap size of 0.25 meV.

**FL-FCI quantum phase transitions**

Figure 1a shows the device image and electrodes used for the measurement of $R_{xx}$ and $R_{xy}$. This device has a moiré wavelength of 11.2 nm and was one of the two devices in which FQAHE was originally discovered[16]. The magnetic-field-symmetrized $R_{xx}$ (see Methods) as a function of carrier density $n$ and $D$ at base temperature is shown in Fig. 1b. Along the white dashed lines that correspond to filling factors $\nu$ = 2/5, 3/7, 4/9, 5/11, 5/9, 4/7, 3/5 and 2/3, FCI states can be seen as local minima of $R_{xx}$ versus $\nu$ (indicated by the orange star for the 3/7 state, for example). For filling factors below 1/2, at higher and lower $D$ values than the range of FCI state, a valley-symmetry-broken FL state and a CI state are observed and labeled by the red triangle and pink hexagon, respectively. The intermediate state between FL and FCI is labeled by the green diamond. Figure 1c-f show the $\rho_{xy} - \rho_{xx}$ plots for states along the dashed lines in Fig. 1b for $\nu$ = 3/7, 4/9, 5/11 and 2/5, respectively. We found that the states between FL and FCI fall on semi-circles at the base temperature, while the states between FCI and CI form a vertical line feature. Note that we converted $R_{xx}$ to $\rho_{xx}$ by using a geometric aspect ratio of around 1.3 (see Methods).

The observation of the semi-circles of resistivities agrees with the spatially separated two-phase picture, in which FL and FCI phases occupy white and grey regions respectively as illustrated in Fig. 1g. Using $\nu$ = 3/7 as an example, when the device area is dominated by one of the two phases under ideal conditions, the macroscopic transport is to follow $(\rho_{xy}, \rho_{xx}) \sim (0, 0)$ and $(7h/3e^2, 0)$ correspondingly. As we tune $D$ by starting from the FL-dominated side, the regions of the FCI phase gradually expand and redistribute the current inside the sample. Along such a process, the net resistivities(conductivities) deviate from the pure FL state, $(\rho_{xy}, \rho_{xx}) \approx (0,0)$, following the semi-circle trajectory determined by the resistivities (conductivities) of the two phases, and eventually reach $(\rho_{xy}, \rho_{xx}) \sim (7h/3e^2, 0)$. At even lower $D$ values, the device gets into an insulating state that corresponds to $\rho_{xy} \sim 7h/3e^2$ and a diverging $\rho_{xx}$. This picture applies to $\nu$ = 3/7, 4/9, 5/11 and 2/5 well as shown in Fig. 1c-e.

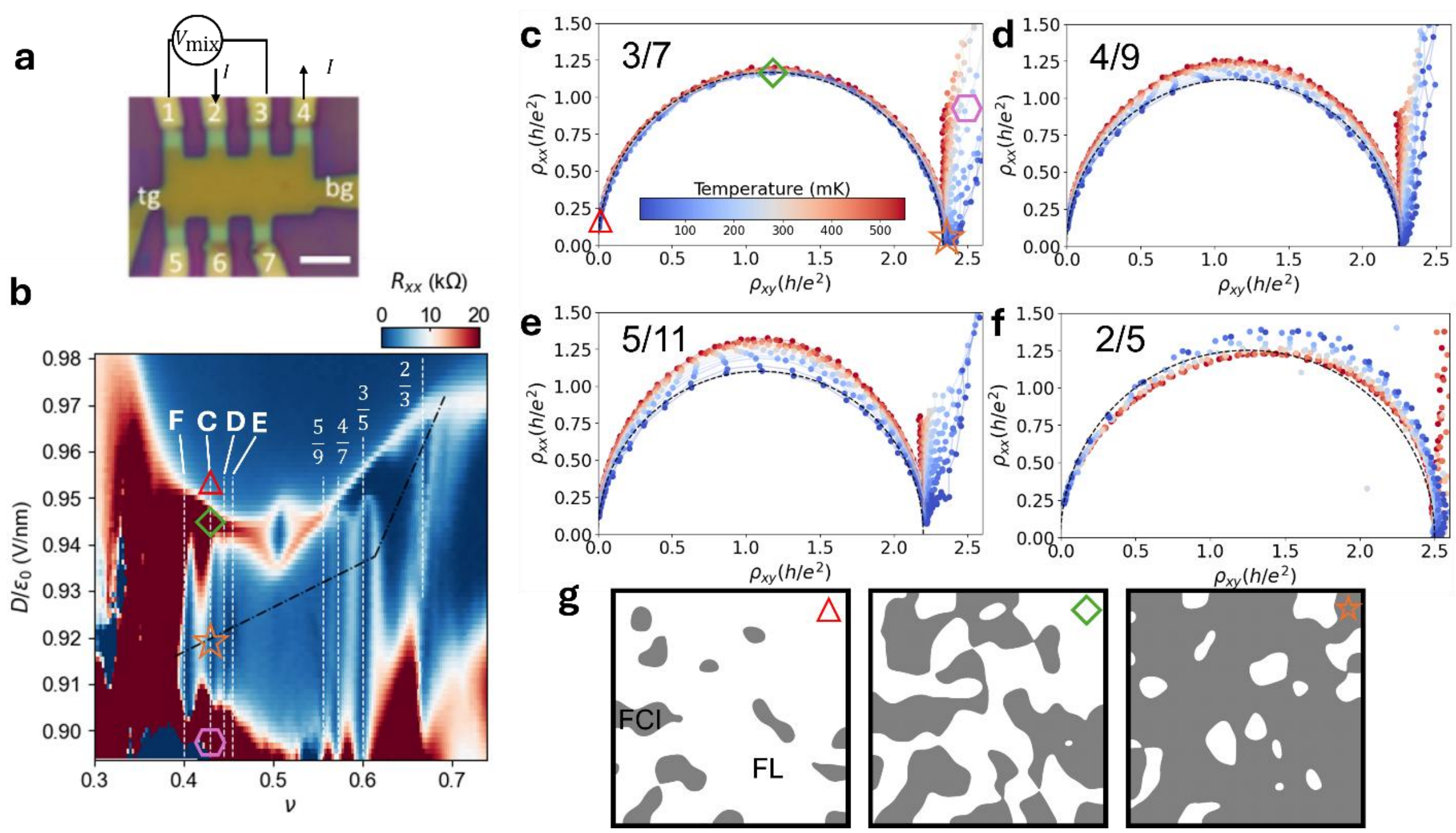

***Fig. 1. Resistivity semi-circles of FL-FCI quantum phase transitions in a pentalayer rhombohedral graphene/hBN moiré superlattice. a**, Optical micrograph of the device reported on in this paper. The probe configuration used to measure $R_{xx}$ and $R_{xy}$ is shown. **b**, Color map of longitudinal resistance $R_{xx}$ as a function of moiré filling factor $\nu$ and D. Vertical stripes corresponding to FCIs can be seen in the intermediate range of D, above (below) which is the Fermi liquid state (the correlated insulating (CI) state). The thermal activation data in Fig. 4 were taken along the dotted-dashed line. **c-f**, Longitudinal resistivity $\rho_{xx}$ versus Hall resistivity $\rho_{xy}$ plots, corresponding to the dashed lines in **b** for $\nu = 3/7,\ 4/9,\ 5/11$ and $2/5$ at varying temperatures. The red triangle, green diamond, orange star and pink hexagon markers in **c** correspond to four (n, D) points in **b** at filling factor $3/7$. These four states are the FL state, the intermediate state between FL and FCI, the FCI state, and the CI state, respectively. The dashed black lines are the predicted perfect semi-circles determined only by $\rho_{xy}$ of the FCI state. At all four fractional fillings, the lowest temperature data corresponding to the FL-FCI quantum phase transition fall on the semi-circle. Extra data forming vertical lines at $\nu = 3/7, 4/9,$ and $5/11$ result from the transition from FCI to the CI state with a diverging resistivity. **g**, Cartoon to illustrate the evolution of a two-phase (shown as white and gray domains) mixture as the phase transition occurs. When the relative fraction of the two phases moves through the three panels of the figure, the resistivity tensor goes through a semi-circle trajectory in the $\rho_{xx} - \rho_{xy}$ plane.*

## FL-EQAH-FCI/CFL phase transitions

At filling factor $\nu \geq 1/2$, additional EQAH states have been found in previous experiments[27] at the base temperature. Figure 2a highlights $\nu$ = 1/2, 4/7 and 3/5 by dashed lines, which trace the phase transition from FL to EQAH to CFL (for 1/2) and FCI (for 4/7 and 3/5). Figure 2b-d show the semi-circle transitions from $(\rho_{xy}, \rho_{xx})$ = (0, 0) to ($h/e^2$, 0) (FL to EQAH) and ($h/e^2$, 0) to ($h/\nu e^2$,0) (EQAH to FCI), with perfect semi-circles shown by the dashed curves. The agreement between the data and ideal semi-circles is not as good as in Fig. 1. At elevated temperatures, the data gradually changes toward the big semi-circle connecting ($h/\nu e^2$, 0) and the origin.

If there are only two phases, the semi-circle law is exact. For scenarios like the ones shown in Fig. 2b-d, we apply the effective medium approximation to understand the effective conductivity of the mixture of four phases[55]. Specifically, for $\nu = 1/2$ the four phases are CFL, EQAH, Fermi liquid, and an insulator. For $\nu = 4/7$ and $\nu = 3/5$ the four phases are FCI, EQAH, Fermi liquid, and an insulator. In general, the resulting effective conductivity/resistivity lies between the largest and smallest semi-circles that connect the four pure phases on the $(\rho_{xy}, \rho_{xx})$ plot. The relative fractions of the phases are determined by their free energies. In the simulation, we assume the free energy of the states has a *D*-field dependence and the compressible states (and FCIs) have additional entropy contribution to their free energies. By tuning the parameters, we can obtain simulation results that look qualitatively similar to the experimental data, as shown in Fig. 2e-g. The general features are found to be insensitive to the values of parameters. Details of the simulation are summarized in Methods.

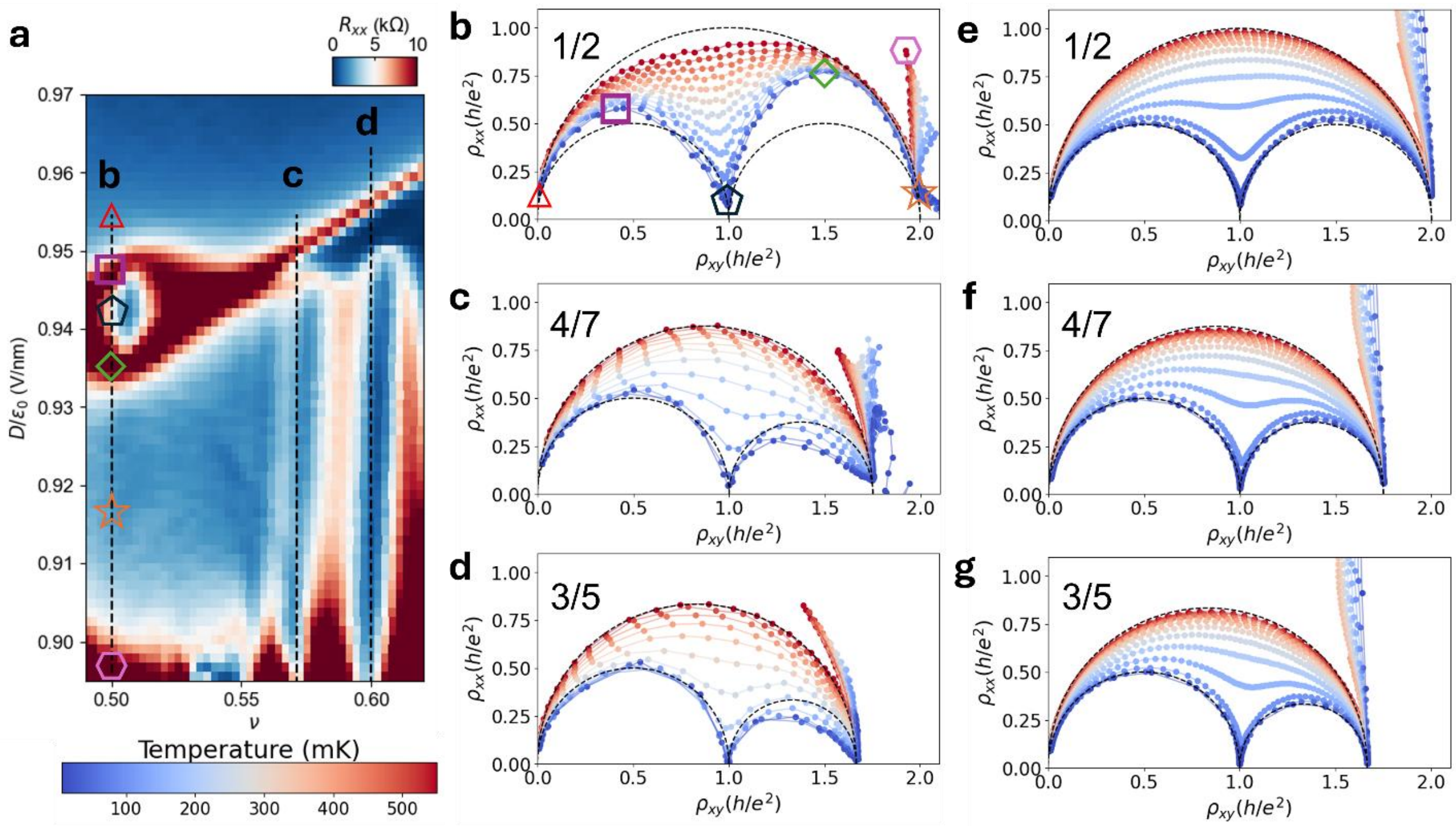

*Fig. 2.* ***Resistivity semi-circles of the FL-EQAH-FCI/CFL phase transitions. a**, $R_{xx}$ colormap showing filling factors at and above $\nu = 1/2$, and dashed lines trace the D-field-driven quantum phase transitions at $\nu = 1/2, 4/7, 3/5$. The red triangle represents a Fermi liquid state, the black pentagon the EQAH state with Chern number C = 1, the orange star represents the CFL state with $\rho_{xy} \approx 2h/e^2$. The purple square represents the intermediate point between the FL and EQAH state, and the green diamond represents the intermediate point between EQAH and CFL.* ***b-d**, Resistivity plots for mixing chamber temperatures of 7 to 550 mK taken along the dashed lines in*

*a, respectively. The $\nu = 1/2$ plot is annotated with markers corresponding to those on the colormap in **a**. Dashed black lines show perfect semi-circles that connect the three ideal phases. Data at the lowest temperatures show qualitative similarity with the semi-circles, although the level of quantitative agreement varies.* ***e-g**, Effective medium approximation simulations of the transitions shown in **b-d**, in which we consider a mixture of four phases.*

**Fractional quantized anomalous Hall insulators and transitions to FCI**

While the phase transitions at high $D$ values can be described by the semi-circles of resistivities, the transition between the insulating state and FCI at lower $D$ values appear to extend vertically in Fig. 1c-f. To understand this FCI-CI phase transition, we plot the $\nu$-$D$ map of $R_{xx}$ as shown in Fig. 3a and highlight two fractional states at $\nu$ = 3/7 and 4/9 by dashed lines. Instead of resistivity plots, Fig. 3b and 3c show the conductivity plots along the dashed lines in Fig. 3a, respectively. We found that the states during the FCI-CI transition fall on a semi-circle defined by $(\sigma_{xy}, \sigma_{xx}) = (\nu e^2/h, 0)$ and (0, 0), corresponding to that of an ideal FCI state and an ideal insulator, respectively. The FCI-FL transition appears to diverge vertically on the $\sigma_{xy}$-$\sigma_{xx}$ plot. We note that the geometric aspect ratios we used to convert $R_{xx}$ to $\rho_{xx}$ and subsequently $\sigma_{xx}$ in Fig. 3 are the same as in Fig. 1 and Fig. 2, ensuring the self-consistency of our data analysis.

The two-dimensionality of the conductivity and resistivity matrices guarantees that if one satisfies a semi-circle law, so does the other[31]. The radii of the two semi-circles may be very different, which makes their observations easier or harder. When the phase transition involves the FL that has diverging longitudinal conductivity (insulator that has diverging longitudinal resistivity), plotting resistivities (conductivities) is the right way to reveal the semi-circle law, as shown in Fig. 1 (Fig. 3). This is so, since, for example, the radius of the semi-circle that involves an insulator is infinite when plotted for resistivities and is finite when plotted for conductivities.

We note that the insulator at low $D$ is substantially different from a trivial insulator. The latter features diverging $\rho_{xx}$ and zero $\rho_{xy}$. In contrast, it has been shown in conventional 2-dimensional electron gas (2DEG) systems that a quantized Hall insulator can exist[65-69], which harbors diverging $\rho_{xx}$ and quantized $\rho_{xy} = h/e^2$. The semicircle relation of conductivities holds as long as $\rho_{xy} = h/e^2$, as noted by Shahar *et al*[76]. We observed almost quantized $\rho_{xy}$ at $\nu$ = 3/7 during the FCI-insulator transition in Fig. 1b and 1c, and deep into the insulator state ($\rho_{xx}$ up to ~6*$\rho_{xy}$, as shown in Fig. 3d). At the same time, the conductivity semicircle is well-behaved as shown in Fig. 3c. Similar observations are made for $\nu$ = 4/9 (see Extended Data Fig. 4). These observations suggest the existence of a new type of insulator, which we name as the fractional quantized anomalous Hall insulator, in analogy to the quantized Hall insulator shown in conventional 2DEG systems.

Figure 3d-f show more details of the quantum phase transition from FCI to fractional quantized anomalous Hall insulator at $\nu$ = 3/7. $\rho_{xx}$ curves at varied temperatures cross each other

at $D_c/\varepsilon_0$ = 0.911 V/nm, as seen in the inset of Fig. 3d. Above this critical value ($D_c$), $\rho_{xx}$ increases with increased temperatures, which is typical for quantum Hall liquid, while below $D_c$ it decreases with increasing temperature, as expected from a quantized Hall insulator. Figure 3e shows the relation between current $I$ and longitudinal voltage drop $V_{xx}$ in the vicinity of $D_c$. At exactly $D_c$, the $I$-$V_{xx}$ curve follows a linear relation that is indicated by the dashed diagonal line. In contrast, nonlinearities with opposite curvatures develop at above and below $D_c$. There is a symmetry of all the curves about the linear $I$-$V_{xx}$ line in Fig. 3e. This symmetry is further exemplified by the overlap between three curves below $D_c$ and three mirror-reflected curves above $D_c$, as shown in Fig. 3f.

Such symmetry of the $I$-$V_{xx}$ curves at around a transition point was observed in GaAs-based 2DEGs (near a critical magnetic field $B_c$), and was interpreted as a signature of charge-flux duality[69,76], which is a consequence of the flux attachment approach to the FQHE. Indeed, flux attachment holds also as an approach for the understanding of the FCI states[30].

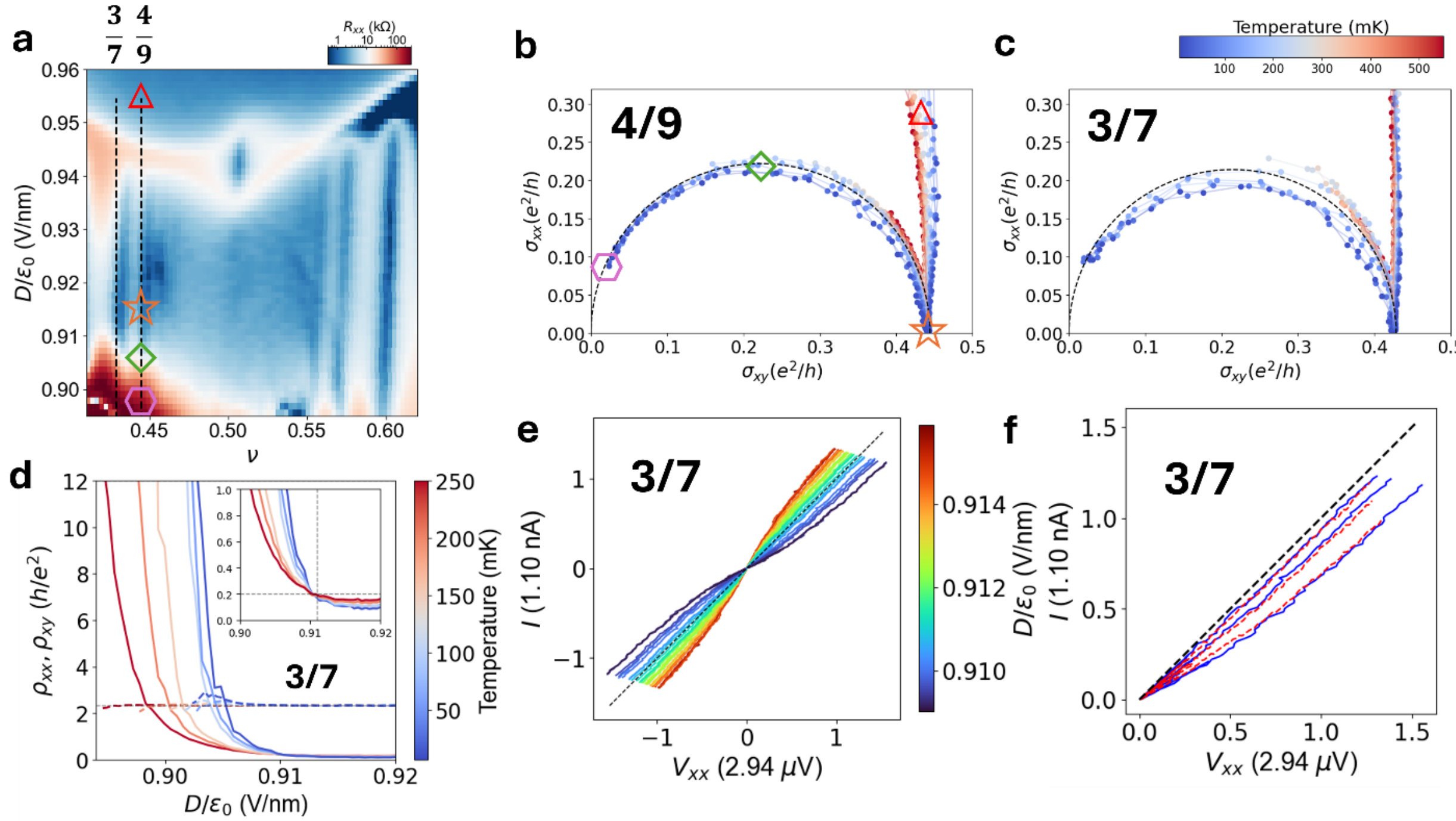

***Fig. 3. Conductivity semi-circles and fractional quantized anomalous Hall insulators. a**, $R_{xx}$ color map, with black dashed lines tracing the D-field-driven quantum phase transitions at $\nu$ = 3/7 and 4/9. At $\nu$ = 4/9. The red triangle, orange star, pink hexagon, and green diamond represent a Fermi liquid, the FCI state, the insulating state, and the intermediate state, respectively. **b**,**c**, Plots of $\sigma_{xx}$ vs. $\sigma_{xy}$ at varying temperatures for 4/9 and 3/7 taken along dashed lines in **a**, respectively. The transitions from the FCI to the insulator agree well with the semi-circles, while the data corresponding to the FCI-FL transition develops into vertical lines with diverging conductivity. **d**, $\rho_{xx}$ (solid curves) and $\rho_{xy}$ (dashed curves) vs. D at $\nu = 3/7$ along part of the dashed line in **a**. $\rho_{xx}$ diverges with $\rho_{xy}$ remaining approximately quantized. Both the conductivity semi-circles and quantized $\rho_{xy}$ suggest a fractional quantized anomalous Hall insulator, in*

*analogy with the integer quantized Hall insulator observed in conventional 2DEGs. The inset zooms in on the $R_{xx}$ curves to show the 'crossing point' at which the resistance changes from increasing to decreasing with increasing temperature. The vertical and horizontal dashed lines mark $D_c$ and the corresponding resistance determined from **e**. **e**, Current I versus voltage-drop $V_{xx}$ at D values near the transition at the base temperature, showing symmetry about the I-$V_{xx}$ curve at a critical value $D_c$, further reminiscent of the quantized Hall insulator in GaAs-based 2DEGs. **f**, Part of data in **e**, but with the curves at D > $D_c$ (red dashed lines) reflected about the diagonal line (dashed black line) to show their symmetry to curves at D < $D_c$ (blue solid lines).*

**Temperature-dependent transport in FCI states**

Figure 4a shows the $R_{xx}$ at zero magnetic field along the black dashed-dotted line in Fig. 1b at varied temperatures. The resistance of the FCI states highlighted by yellow color shows a clear increase as the temperature increases. The resistance peak between these shaded FCI states, however, remains roughly unchanged as the temperature is increased (except for between $\nu$ = 3/5 and 2/3, where the change is due to the disappearance of the EQAH state). In contrast, the FCI states closer to half-filling and the resistance peak between them show similar increases with a temperature increase. The change of resistance at and between these weaker FCI states is similar to that of the likely CFL state at half-filling. Such observations are also found when a 2 T out-of-plane magnetic field is applied, as shown in Fig. 4b. Strictly speaking, in the presence of a non-zero magnetic field, FCIs are shifted from the charge densities corresponding to fractional moiré filling factor $\nu$ to $\nu^* = n/(\frac{B}{\Phi_0} + n_m)$, where $n$ and $n_m$ are the densities of electrons and moiré sites respectively, to account for the shift of FCI states due to the Středa relation[63,64].

The two types of temperature-dependence we observed suggest qualitatively different physical pictures: in the first scenario, the change of resistance at fractional fillings is dominated by the thermal activation of the insulating bulk of the FCI states; in the second scenario, there is an increase of resistance even for the likely-CFL, which is gapless (no plateau of $R_{xy}$ or dip in $R_{xx}$ have been observed at $\nu$ = 1/2), and this resistance increase dominates the behavior in a wide range of filling factors including those of the un-shaded FCI states.

Figure 4c&d examine these two pictures quantitatively. Using $\nu$ = 2/5 as an example of the former, we compare the temperature-dependent $R_{xx}$ to two models: the first model uses the combination of a variable-range-hopping term[51-54] ($\exp(-\sqrt{T_0/T})$) and a thermal activation across a fixed gap term[39-47] ($\exp(-\Delta/2k_BT)$); the second model uses the combination of a $T^2$ term and a constant term[28]. By optimizing the parameters, we found a reasonable agreement between our data and the first model, while the second model always deviates noticeably from our data regardless of the parameters. In contrast, $R_{xx}$ at $\nu$ = 1/2 can be well-described by the second model, as shown in Fig. 4d. This is similar to the temperature-dependence in a Fermi liquid and agrees with the gapless nature of the likely CFL state at half-filling of this moiré system. Although the fitting for 1/2 with $A\exp(-\Delta/2k_BT) + B\exp\left(-\sqrt{T_0/T}\right)$ seems reasonable, the range of resistance change in Fig. 4d is much smaller than in Fig. 4c, making the fitting much less meaningful.

Applying the same fitting procedure to other FCI states that are shaded in Fig. 4a, we try to extract the thermal activation gap Δ in the first model, as shown in Fig. 4e. The gap of 2/3 being larger than 3/5 is consistent with the 'V'-shape of gap versus $(\nu - 1/2)$ that has been observed in the FQH systems and is predicted by composite fermion theory[28,56,57]. We note that the values of Δ, reaching ~0.25 meV (~ 3 K) for the most robust FCI states, are comparable to those observed in the FQH states in conventional 2DEGs[39-47] but smaller than FQH states in graphene at similar charge densities[50,58]. One possible explanation of the smaller gap size compared to FQH states in graphene with similar charge density is the difference in dielectric constant between monolayer/bilayer graphene and pentalayer graphene. Density functional theory simulations suggest that the dielectric constant should increase with layer number for rhombohedral stacking[59]: bilayer graphene has a dielectric constant of < 3 while pentalayer is more than 8. Although the effective dielectric constant felt by electrons in graphene is also contributed by that of hBN[60], we suspect the self-screening by the pentalayer graphene itself might reduce the FCI gaps compared to FQH gaps in bilayer graphene.

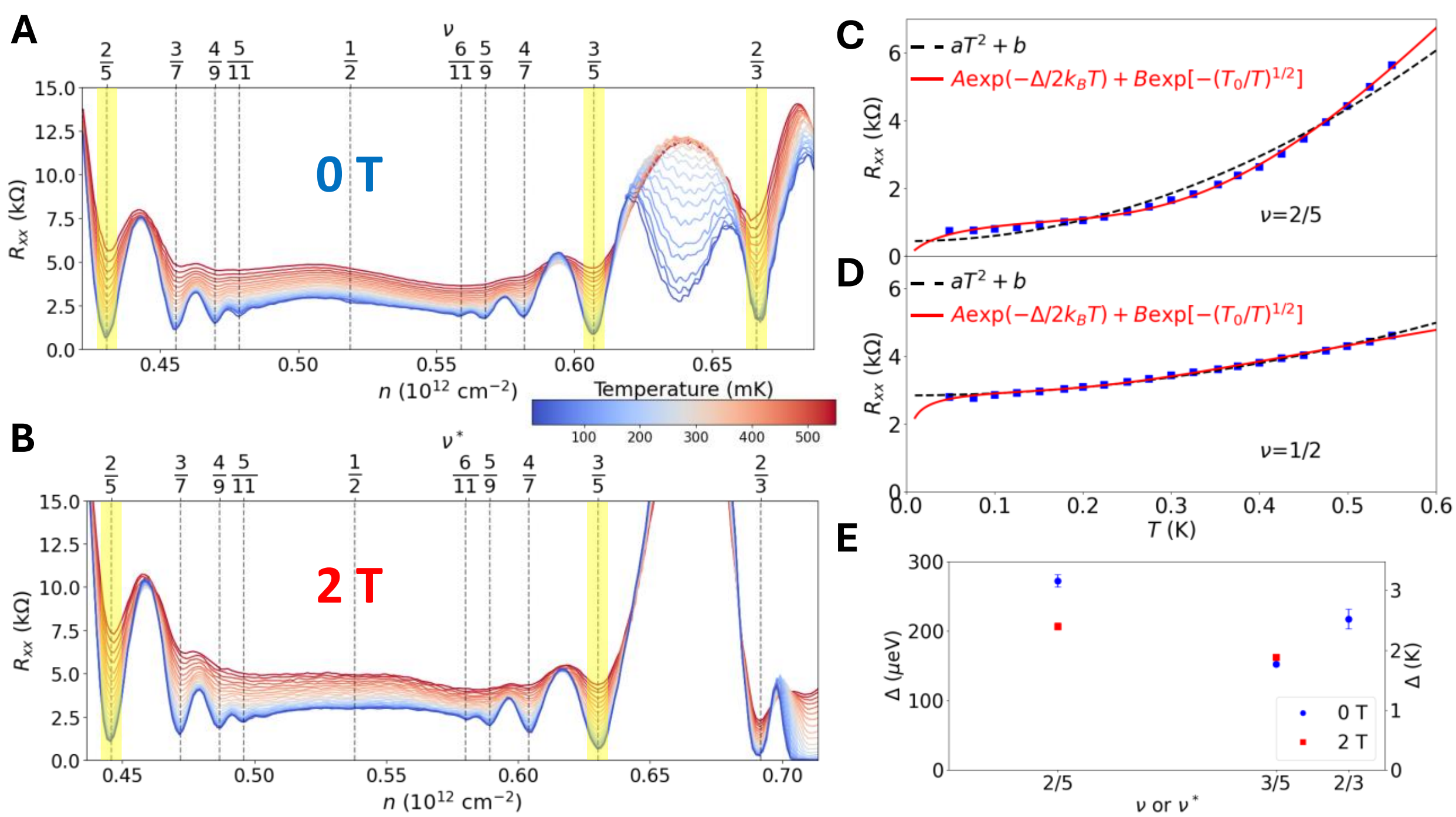

***Fig. 4. Comparison with thermal activation model and transport gaps of FCI states. a,b,*** *Temperature dependence of $R_{xx}$ vs electron density along the dashed-dotted line in Fig. 1a, taken at 0 T and 2 T out-of-plane magnetic fields. The upper x-axis is the moiré filling factor.* ***b****, Similar to* ***a****, but at 2 T, and along a different trajectory in $(n, D)$ space (see Extended Data Fig. 5). The data are qualitatively similar at these two magnetic fields, except for the suppression of the EQAH*

*state by magnetic field. The upper axis is the modified filling factor $\nu^*$.* ***c****, Plot of $R_{xx}$ at $\nu$ = 2/5. The dashed black line is a fitting of experimental data to a $T^2$ law, and the red solid line is a fit to the sum of a thermally activated resistance term and a variable range hopping term.* ***d****, The same as* ***c****, but for filling fraction $\nu = 1/2$.* ***e****, The gap energies extracted from thermal activation at 0 T and 2 T. Only the most robust FCI states whose thermal activation behaviors are not strongly convoluted by the $T^2$ behavior are shown, which are highlighted by yellow shades in* ***a*** *and* ***b****.*

Our data and analysis focus on a new regime of FQAHE/FCI physics in RG/hBN moiré superlattices, with the lowest electron temperature and highest device quality to date. They provide important insights into the nature of FCIs and their associated quantum phase transitions. The 11 fractional states we observed all happen at filling factors in the Jain sequence of two-flux CFL[28, 29], indicating a high similarity to the FQH system and suggesting an underlying flat Chern band that mimics the Landau level. The observation of similar temperature-dependent transport properties in the three most robust FCI states also mimic those in the conventional FQHE systems. These observations should help in resolving the microscopic mechanism of FQAHE/FCIs in the RG/hBN system, which has been under debates and remained unclear theoretically[20-26]. The observed semi-circles provide the first experimental evidence of the microscopic two-phase picture for the phase transitions between FCI and other *D*-field-tuned quantum phases, which was previously demonstrated only by tuning magnetic field in conventional 2DEGs for the IQHE[33,34] and for the $\nu = 1/3$ FQHE[36].

The observed temperature-dependent transport properties provide the necessary background of further investigations into anyon charge and statistics through interferometry[70,71], direct microscopic imaging through scanning-probe techniques[72,73], and enhancement of device quality by reducing disorders during the fabrication process[74,75]. By clarifying the microscopic details of the *D*-field-induced quantum phase transitions and revealing the fractional quantized anomalous Hall insulators, our work also paves the way to further engineering gate-defined junctions. For example, one can explore parafermions at the gate-defined FQAHE/superconductivity interface—these two phases have been observed to coexist in close proximity on the *n-D* diagram of RG/hBN devices[2].

**Acknowledgements**

We acknowledge helpful discussions with S.D. Sarma, S. Todadri, and R. Ashoori. This study was supported by the US Department of Energy, Office of Science, Basic Energy Sciences, Materials Sciences and Engineering Division under grant no. DE-SC0025325. Z.H. acknowledges support from a fellowship from the Physics Department of MIT. Device fabrication was performed at the Harvard Center for Nanoscale Systems and MIT.nano. T.H. acknowledges support from a Mathworks Fellowship. K.W. and T.T. acknowledge support from the JSPS KAKENHI (grant nos. 20H00354, 21H05233 and 23H02052) and the World Premier International Research Center Initiative, MEXT, Japan. A.S. acknowledges support from the ISF, ISF Quantum Science and Technology (grant 2074/19) and the DFG (CRC/Transregio 183).

**Author Contributions**

L.J. supervised the project. Z.H. and Z.L. performed the DC magneto-transport measurement. T.H. and Y.Y. fabricated the devices. S.Y., Z.H., J.Y., J.S. and T.H. helped with installing and testing the dilution refrigerator. X.Y. and A.S. performed the simulation. K.W. and T.T. grew hBN crystals. All authors discussed the results and wrote the paper.

**Competing Interests** The authors declare no competing interests.

**Data and materials availability**

All data that supports the findings of this study are present in the main text or extended data and available from the corresponding authors upon request.

## Methods

### Device fabrication

The graphene and hBN flakes were prepared by mechanical exfoliation onto $SiO_2$/Si substrates. The rhombohedral domains of pentalayer graphene were identified and confirmed using IR camera, near-field infrared microscopy, and Raman spectroscopy and isolated by cutting with a femtosecond laser. The van der Waals heterostructure was made following a dry transfer procedure. We picked up the top hBN, graphite, middle hBN, pentalayer graphene using polypropylene carbonate film and landed it on a prepared bottom stack consisting of an hBN and graphite bottom gate. We aligned the long straight edge of the graphene and hBN flakes to form a large moiré superlattice. The device was then etched into a multiterminal structure using standard e-beam lithography and reactive-ion etching. We deposited Cr–Au for electrical connections to the source, drain and gate electrodes.

### Transport measurement and (anti-)symmetrization procedures

The device was measured mainly in a Bluefors LD250 dilution refrigerator at MIT with a lowest electronic temperature of around 40 mK. Stanford Research Systems SR830 and SR860 lock-in amplifiers and SP1004 voltage preamplifiers from Basel Precision Instruments were used to measure the longitudinal and Hall resistance $R_{xx}$ and $R_{xy}$ with an AC frequency at 17.77 Hz. The AC currents are generated by the SR860 lock-in through a 10 MΩ resistor. The AC current excitation was limited to be 1 nA or below. Basel Precision Instruments preamps were used to measure differential currents and voltages. Keithley 2400 source-meters were used to apply top and bottom gate voltages. Top-gate voltage $V_t$ and bottom-gate voltage $V_b$ are swept to adjust doping density $n = (C_t V_t + C_b V_b)/e$ and displacement field $D/\varepsilon_0 = (C_t V_t - C_b V_b)/2$, where $C_t$ and $C_b$ are top-gate and bottom-gate capacitance per area calculated from the Landau fan diagram.

The device is a modified Hall bar, such that we measure a resistance $R_{\text{mix}}$ which contains both $R_{xx}$ and $R_{xy}$ components. $R_{xx}$ ($R_{xy}$) is extracted from symmetrizing (anti-symmetrizing) the mixed resistance signal with respect to positive and negative magnetic field; $R_{xx} = \big(R_{\text{mix}}(+B) + R_{\text{mix}}(-B)\big)/2$ and $R_{xy} = \big(R_{\text{mix}}(+B) - R_{\text{mix}}(-B)\big)/2$ . $\pm 0.1$ T is used for this procedure unless otherwise stated.

### Extraction of resistivities and conductivities

The measured quantities are $R_{xx}$ and $R_{xy}$, but the semi-circle law describes the relation between $\rho_{xx}$ and $\rho_{xy}$. In our modified Hall bar geometry, $\rho_{xy} = R_{xy}$, however, while $R_{xx}$ is proportional to $\rho_{xx}$, $R_{xx}$ depends on geometrical details so we expect that $\rho_{xx} = cR_{xx}$ for some geometrical scaling factor $c$. For a conventional Hall bar, this scaling factor is easily determined by the aspect ratio of the Hall bar. In our case for the modified Hall bar geometry, calculating this geometrical factor is less straightforward, but we did simulations using COMSOL Multiphysics® and the Electric Currents interface of the AC/DC module[61,62] and determined it as 1.32. This aspect ratio results in a good agreement between experimental data and the semicircle at $\nu$= 3/7. But for other fractional fillings, we found that a different aspect ratio is needed to align the semicircles with the corresponding data. This aspect ratio is always within 30% of 1.32. For any given fraction, we apply the same scaling factor for each temperature. The conductivity is computed from the resistances using this same scaling factor such that, for any fraction, $\sigma_{xx} = \rho_{xx}/\left((\rho_{xx})^2 + \left(\rho_{xy}\right)^2\right)$ and $\sigma_{xy} = \rho_{xy}/\left((\rho_{xx})^2 + \left(\rho_{xy}\right)^2\right)$.

**Improvement of residual resistance of FCI states**

The device shown in the main text was originally reported[16,27] as device 2 in an earlier manuscript where the residual $R_{xx}$ of the fractional states was high, on the order of several kΩ and with improved electronic filtering diminished to around 1 kΩ. The device has been thermally cycled several times since it was first measured, and we have found that these thermal cycles have enhanced the FQAH states. For the most robust fractions such as 3/7 and 3/5, the $R_{xx}$ residue has been reduced to ~500 Ω, well below that of device 1 in the original two manuscripts.

**Electron temperature**

Near the base temperature of 7 mK, the electrons are not fully thermalized with the mixing chamber. We observe that $R_{xx}$ does not appreciably change below a mixing chamber temperature of 50 mK, so we assume that the electron temperature is thermalized with the mixing chamber temperature at 50 mK and above. Accordingly, we only fit data taken at 50 mK or warmer.

**Effective medium approximation simulations**

We use the effective medium approximation (EMA) to model transport in an inhomogeneous system composed of multiple "puddles" (domains) of different phases, each characterized by a conductivity tensor $\hat{\sigma_i}$. EMA replaces the heterogeneous mixture by a homogeneous effective medium with conductivity tensor $\hat{\sigma}$.

In the EMA picture, a puddle of phase $i$ is embedded in the effective medium. Because $\hat{\sigma_i} \neq \hat{\sigma}$ in general, the local electric field and current inside the puddle differ from those in the host, producing an induced current $\Delta I_i$. The effective tensor is obtained from a mean-field self-consistency condition: the induced current averages to zero when weighted by the phase fractions.

$$\sum_i \delta_i \Delta I_i = 0$$

Here, $\delta_i$ is the areal fraction of phase $i$, with $\sum_i \delta_i = 1$.

In two dimensions, EMA is known to perform particularly well for two reasons: it captures the correct percolation threshold, and for a two-phase mixture, it is consistent with the semicircle law which is exact for a two-phase mixture.

More precisely, $\Delta I_i \propto \left(\sigma_d \mathbb{1} - \frac{1}{2}\hat{\sigma} + \frac{1}{2}\hat{\sigma_i}\right)^{-1} (\hat{\sigma} - \hat{\sigma_i})$, where $\mathbb{1}$ is the 2×2 identity matrix and $\sigma_d$ denotes the longitudinal part of the effective tensor $\hat{\sigma}$. So, one can obtain the effective tensor by solving:

$$\sum_i \delta_i \left(\sigma_d \mathbb{1} - \frac{1}{2}\hat{\sigma} + \frac{1}{2}\hat{\sigma_i}\right)^{-1} (\hat{\sigma} - \hat{\sigma_i}) = 0$$

To compare with experiment, we assume the system may contain puddles of several phases: CFL, FL, FCI, extended quantum Hall state, and trivial insulator whose populations depend on the temperature and $D$ via the Boltzmann distribution $\delta_i \propto e^{-F_i/k_B T}$ . The possible phases at a specific density are determined by the conductivities and resistivities that are observed as $D$ is varied, keeping the density constant. In order to determine $\delta_i$ we model the free energy $F = E - TS$ of phase $i$ as a function of temperature and $D$ by:

$$F = \alpha(D - D_i)^2 - s_i T^2$$

Here, $D_i$ is the value of $D$ at which a phase $i$ has a minimum of energy, and $s_i$ controls the strength of the entropy contribution. For FL phase, the entropy is known to be linear in $T$, while for CFL it is proportional to $T \log T$ for long range coulomb interaction. In practice, we approximate both of them to be linear in $T$ with the same coefficient. And for FCI phases we assume the same entropy dependence as the longitudinal resistivity is of the same order which implies similar amount of dissipation channels. For the insulator and the extended quantum Hall state, we assume the entropy contribution is much smaller and set the corresponding $s_i$ to 0 in practice. We tune the values of $\alpha$ and of $s_i$ (which is either 0 or a constant $s$ depending on the corresponding phase) to match the experimental data. We find the general feature is insensitive to the parameters. From the expression for the free energy of different phases, we could determine the relative fractions of the phases and therefore the effective conductivity from effective medium approximation.

Along the axis of $D$, at the densities we are interested in, there are usually three transitions (for example, between a FL, an EQH, an FCI/CFL, and an insulator). When these transitions are well separated at low temperature, the system at any $D$ could be thought of being composed of two phases, and the resistivity and conductivity show a semi-circle independent of the dependence of $\delta_i$ on $D$, i.e., independent of the values of the free energy parameters. When the system includes three (or more) different phases these values affect the shape of the $\rho_{xx}, \rho_{xy}$ trajectories that the system takes along different paths in the *D, T* plane. We find that the $\rho_{xx}, \rho_{xy}$ trajectories always lie within the maximum region enclosed by the set of semicircles determined by all pairs of phases.

**Determination of the geometric aspect ratio of the device**

Using COMSOL Multiphysics® and the Electric Currents interface of the AC/DC module, the device was simulated in order to determine the geometric aspect ratio of the device, i.e. the ratio between resistivity and resistance[61,62]. We modeled the device in COMSOL using the measured values of the relevant lengths such as channel width, length, and so on. We assumed that $\rho_{xx}$ and $|\rho_{xy}|$ of the rhombohedral graphene layer were close to representative values along the semi-circles. The boundary conditions were: uniform current density of 1 nA entering the source contact, floating potential for the two voltage probe contacts, and grounding the drain contact. To simulate our measurement including anti-symmetrization and symmetrization, we compute the simulated voltage drop across the voltage probes shown in Fig. 1a at a chosen value of $\rho_{xy}$. We then simulate again using $-\rho_{xy}$, modeling the change in sign of the magnetic field. The average of these two voltage drops from the two opposite polarities then simulates $V_{xx}$ and thus $R_{xx}$, whose ratio with the simulation's chosen $\rho_{xx}$ provides an estimate of the device's aspect ratio. These simulations suggest that $\rho_{xx} \approx 1.32 R_{xx}$, across all tested values of $\rho_{xx}$ and $\rho_{xy}$. This aspect ratio of 1.32 is close to those used to scale the semicircles to their theoretical predictions; however, the simulations were unable to explain the need for an aspect ratio which depends on the filling. A 3D plot of the electric potential and current density from an example simulation is shown in Extended Data Fig. 1 where we chose $\rho_{xy} = 25$ kΩ and $\rho_{xx} = 10$ kΩ.

**The spectrum of FCIs**

We observe in total eleven FCIs. In addition to the eight previously reported, we observe Jain states at filling factors at 6/11 and 6/13, as shown in Extended Data Fig. 3. Upon raising the temperature to 300 mK, we also observe the FQAHE at 1/3 filling, see Extended Data Fig. 2. This 1/3 FCI was not recognized earlier because previously the contact was too poor to resolve the suppression of $R_{xx}$ and quantization of $R_{xy}$, because the neighboring regions become very insulating, likely due to the Wigner crystallization. However, the improvement of device quality from thermal cycling allowed us to see the 1/3 FQAHE at an elevated temperature. Because we only observe the 1/3 state at 300 mK, there is an insufficient range of temperatures to analyze its temperature dependence, so we omit its analysis in this work.

**Fractional quantized anomalous Hall insulator at $\nu$ = 4/9**

As shown in Extended Data Fig. 4, similarly to $\nu = 3/7$, we observe a fractional quantized anomalous Hall insulator at $\nu = 4/9$. Extended Data Fig. 4a shows diverging $\rho_{xx}$ accompanied by approximately quantized $\rho_{xy}$, a signature of a Hall insulator, as well as the crossing point of $\rho_{xx}$ versus D where its temperature dependence changes from metallic to insulator-like. Extended Data Fig. 4b and 4c show the symmetry of the *I*-$V_{xx}$ data close to the critical *D* value, further reminiscent of the quantum Hall insulator observed in GaAs heterostructures. The *D* value of the crossing point in the inset of Extended Data Fig. 4a is slightly different the critical $D$ obtained from Extended Data Fig. 4b, possibly due to a heating effect which disfavors the insulating state.

**The line-cut at a 2 T magnetic field**

The linecut used in Fig. 4b is different than that of Fig. 4a in the main text. This is because it is not possible to use the same linecut for both 0 and 2 T for all the FCIs. Upon increasing the magnetic field, the sequence of FCI states shifts significantly in $(n, D)$. Accordingly, to measure the thermal activation at 2 T a different cut is needed than for zero field. A colormap of $R_{xx}$ at +2 T is shown in Extended Data Fig. 5 along with a dashed line showing the cut used for the data in Fig. 4b in the main text.

**$R_{xx}$ Temperature dependence fitting**

From Fig. 4a and 4b, it can be seen that resistance increases significantly with temperature for all densities, not only at the fractional fillings. Near half filling, the temperature dependence is expected to be a metallic $T^2$ dependence. For the weak fractions near half filling, the thermal activation appears to be strongly convoluted by the metallic temperature dependence. On the other hand, the more robust FCIs far away from half filling have temperature dependences much more reminiscent of thermal activation of a transport gap. We fit all the fractions using both methods and compare which fits each fraction best. We find, at 0 T, that only 2/5, 3/5, and 2/3 FCIs are well described by thermal activation plus hopping conduction. As these are the only fractions where the $R^2$ value for the thermal activation plus hopping fits exceeds that of $T^2$ fits, see Extended Data Table 1. Furthermore, these are the only fractions for which the neighboring peaks in resistances do not change significantly for temperature. The other fractions closer to half filling have metallic temperature dependence. As such, we only estimate the gap for the three strongest fractions. For 2 T, a similar conclusion is reached based on the nearby peaks, however, for all fractions except $\nu^* = 2/3$ at 2 T, the thermal activation fits the data better according to the $R^2$ value, see Extended Data Table 2. 2/3 is exceptionally strengthened by the magnetic field, so the change in resistance is likely too small for a proper fit to thermally activated behavior, so for 2 T we only estimate gap sizes for $\nu^* = 2/5$ and $3/5$.

In general, variable range hopping conduction has the form of

$$R_0 \exp\left(-\left(\frac{T_0}{T}\right)^{\beta}\right).$$

For the thermal activation plus hopping conduction fits, we use a variable range hopping term of the Efros-Shklovskii form[52] where $\beta = 1/2$. However, using a term of the Mott form[51] $\beta = 1/4$ fits the data very similarly and does not qualitatively affect our analysis. The fits for all 9 fractions in Fig. 4 are shown in Extended Data Figs. 6-9 using both linear scale and Arrhenius plots.

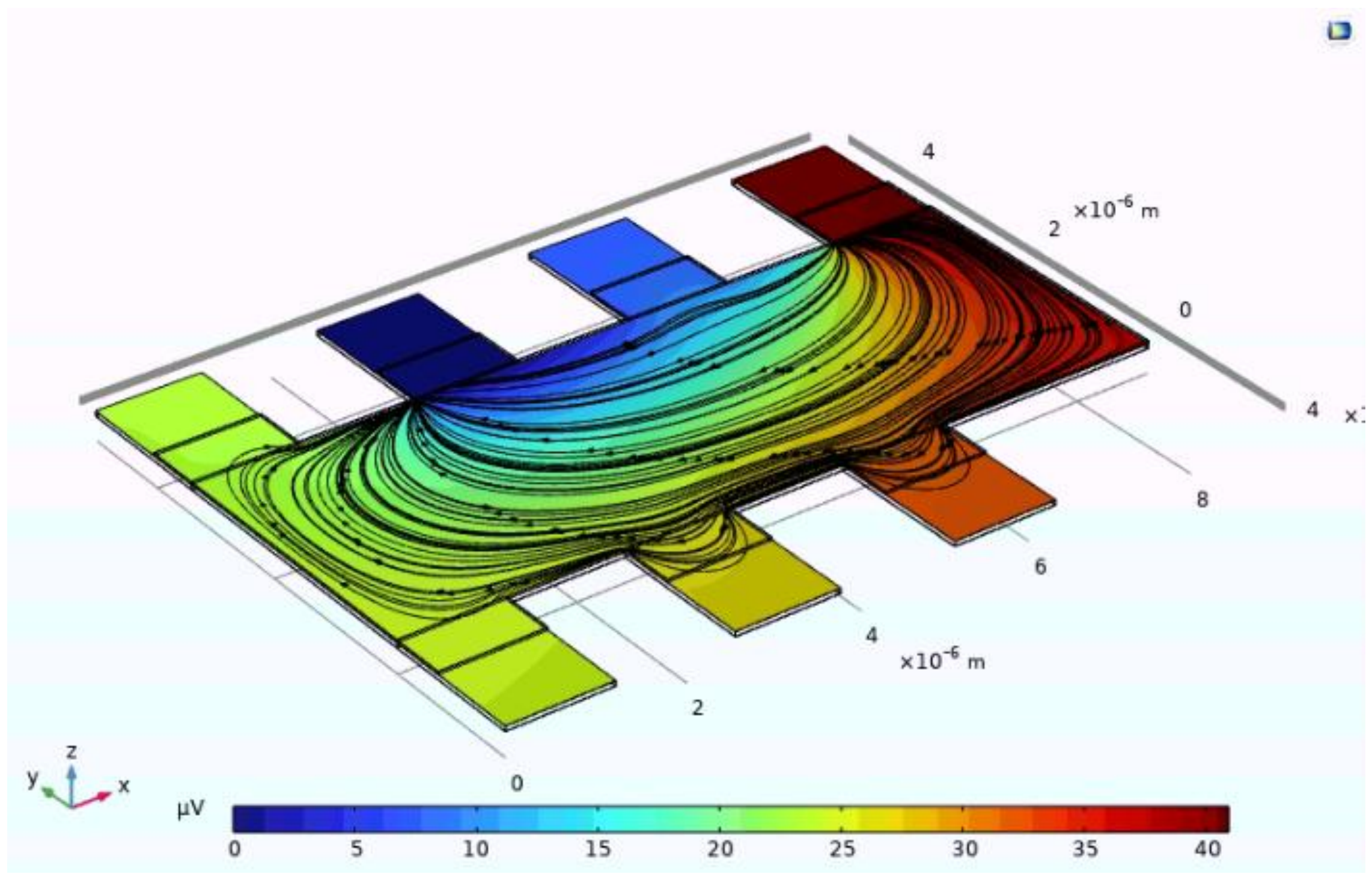

***Extended Data Figure 1*. *Example COMSOL® simulation results for the electric potential and current density within the device.*** *The colormap shows the electric potential through the device, and the black lines with arrows are streamlines representing the current density. From left to right on the side of the device with four contacts, the first and third are floating potential voltage probes. The second and fourth are the drain and source, respectively. Here* $\rho_{xy} = 25$ kΩ *and* $\rho_{xx} = 10$ kΩ.

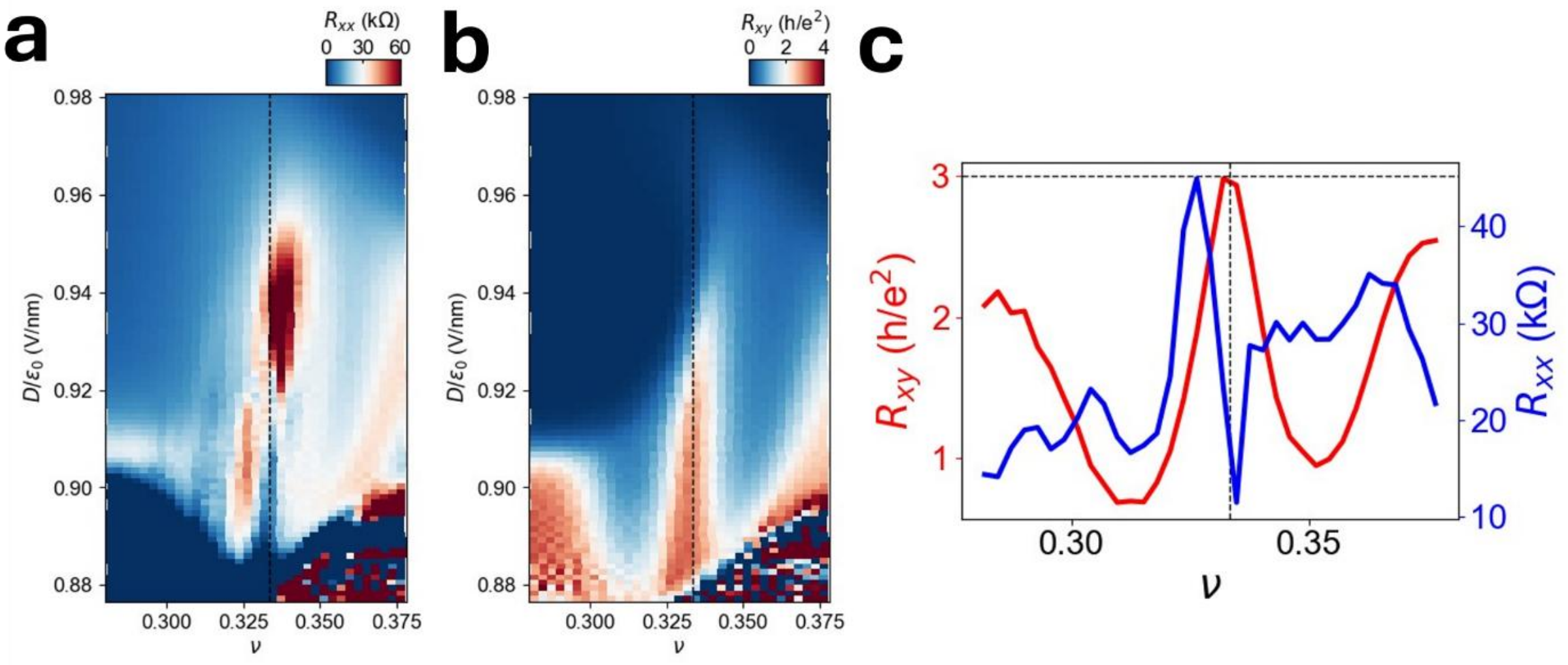

***Extended Data Figure 2. Observation of the fractional quantum anomalous Hall effect at*** $\boldsymbol{\nu = 1/3}$***.*** ***a**,**b***, *$R_{xx}$ and $R_{xy}$ colormaps near filling factor 1/3 at 300 mK suggesting a* $\nu = 1/3$ *FCI*

*state.* ***c****, Linecuts of* $R_{xx}$ *and* $R_{xy}$ *versus* $\nu$ *at* $D/\varepsilon_0 = 0.905$ V/nm*, showing quantization of the latter and a slight dip in the former.*

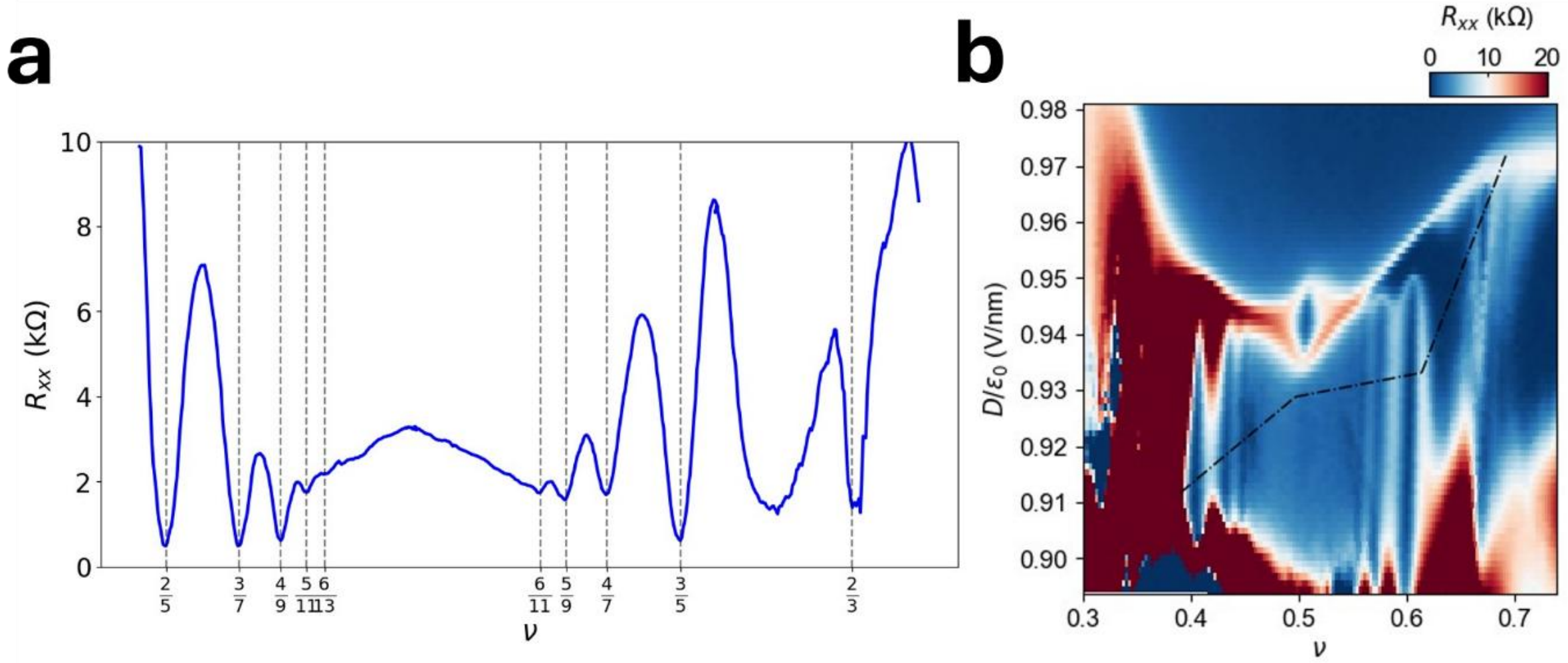

***Extended Data Figure 3. Sequence of Jain states at the base temperature. a****,* $R_{xx}$ *linecut versus moiré lattice filling. We clearly observe 9 fractional states (2/5, 3/7, 4/9, 5/11, 6/11, 5/9, 4/7, 3/5, and 2/3) and see a small dip at ν = 6/13.* ***b****,* $R_{xx}$ *colormap with a dashed-dotted black line showing the trajectory of the cut shown in* ***a****.*

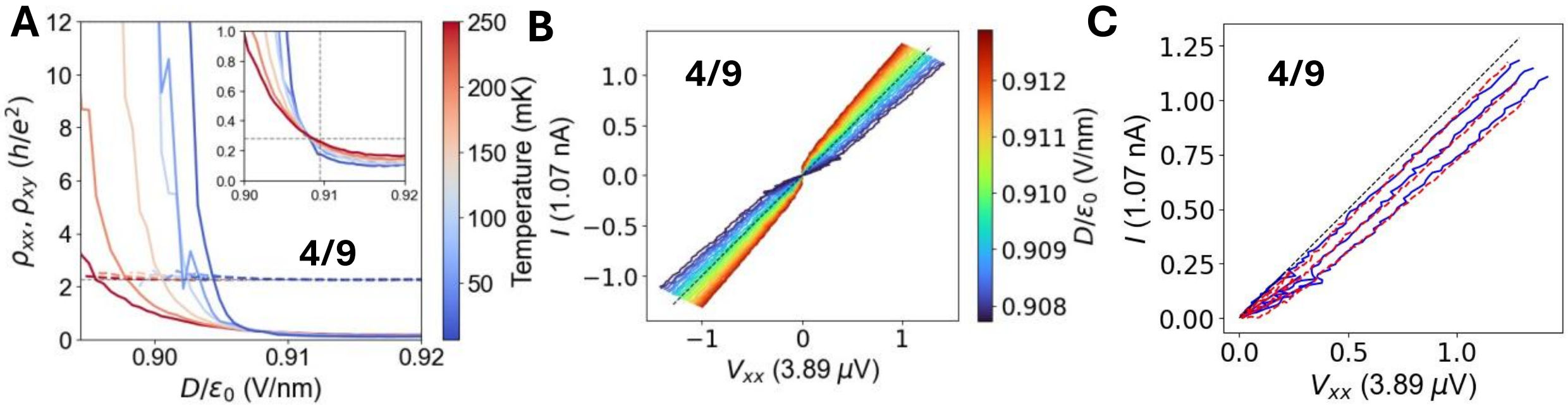

***Extended Data Figure 4. Fractional quantized anomalous Hall insulator at ν = 4/9. a****,* $\rho_{xx}$ *and* $\rho_{xy}$ *vs. D at* ν = 4/9 *along the dashed line in Fig. 3a in the main text.* $\rho_{xx}$ *diverges with* $\rho_{xy}$ *remaining approximately quantized, suggesting observation of a fractionally quantized Hall insulator. The inset is the same* $\rho_{xx}$ *data zoomed in to show the crossing point from metallic to insulating temperature dependence.* ***b****, Current versus* $V_{xx}$ *at D close to the transition between FCI and the quantized Hall insulator at* ν = 4/9*, showing symmetry about the I-V curve at a critical D, further reminiscent of the quantized Hall insulator. The vertical and horizontal dashed lines in the inset of* ***a*** *mark the critical D and its corresponding* $\rho_{xx}$ *as determined by the I-V curves.* ***c****, The same data as in* ***b****, but with the curves at higher D than the critical point (red dashed lines) reflected across the critical line (dashed black line) to emphasize their symmetry to curves at lower D than the critical point (blue solid lines).*

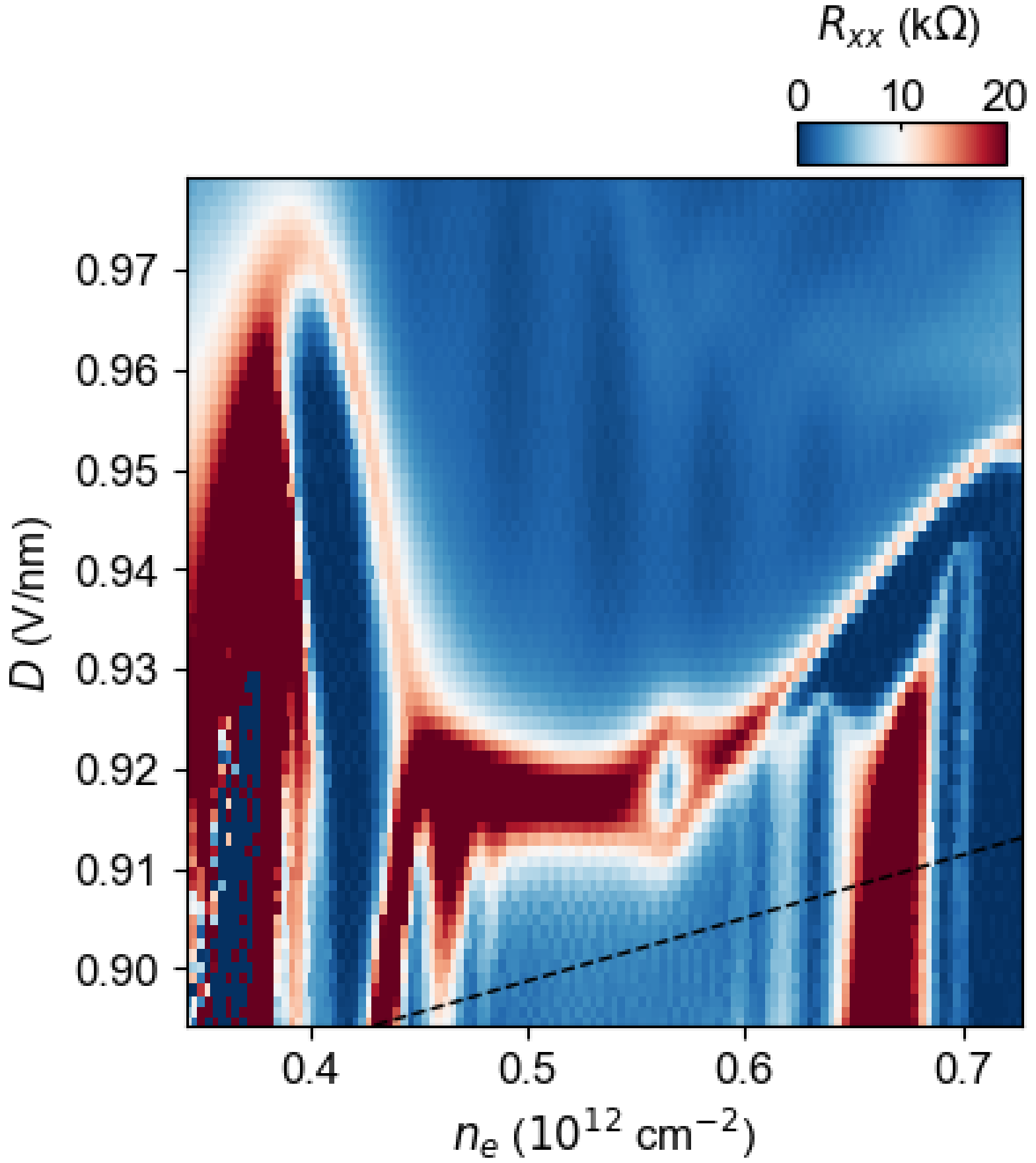

***Extended Data Figure 5. $R_{xx}$ map at +2 T.*** *$R_{xx}$ colormap of the FCI region measured at 2 T. The dashed black line shows the trajectory along which the 2 T data are taken in Fig. 4 of the main text.*

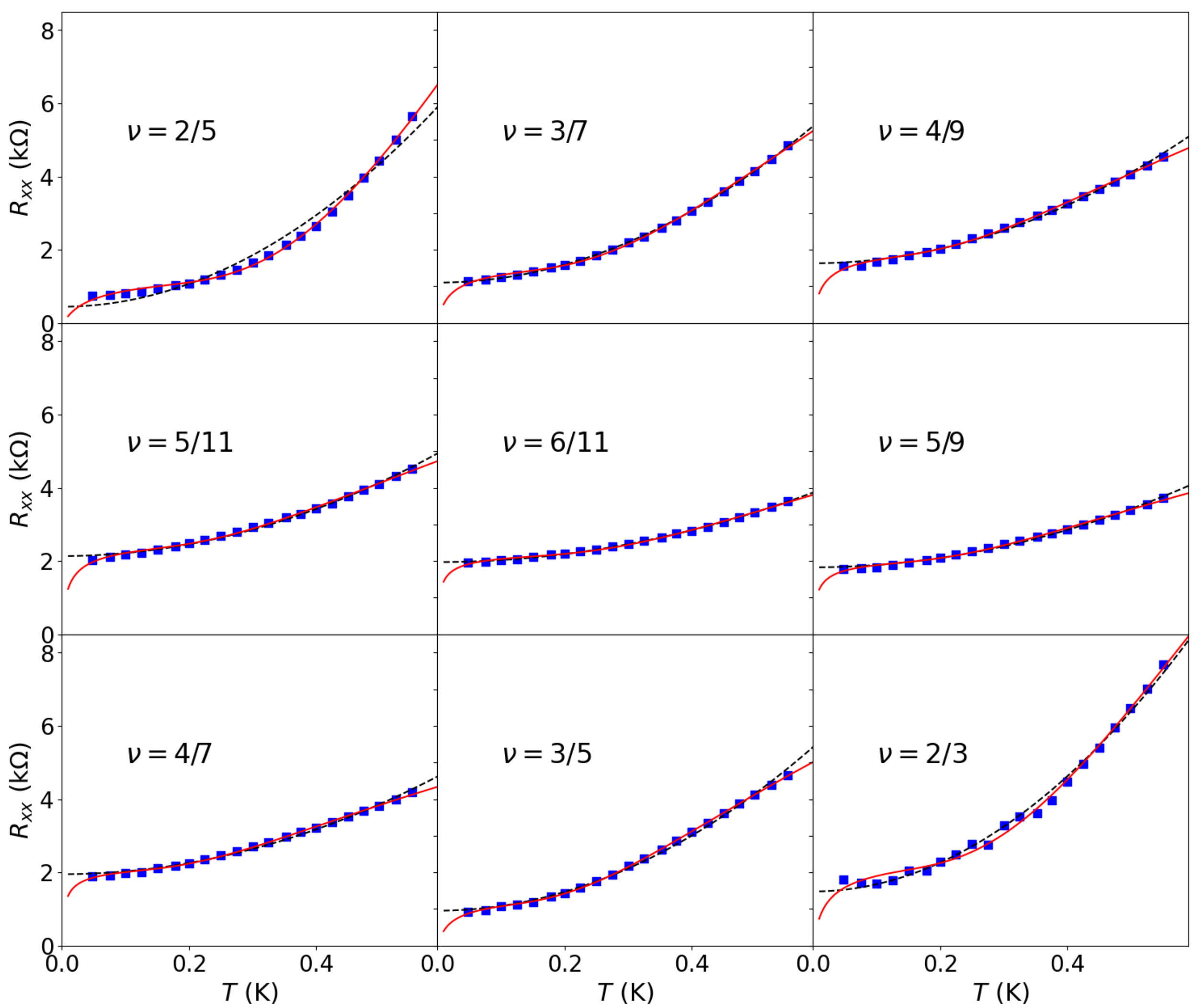

***Extended Data Figure 6. Comparison of thermally activated vs. metallic temperature dependence fits at zero field.*** *$R_{xx}$ versus $T$ plotted on a linear scale for each FCI state at zero field. The dashed black lines are fits to a $T^2$ law and the solid red lines are fits to a sum of thermal activation plus Efros-Shklovskii hopping conduction.*

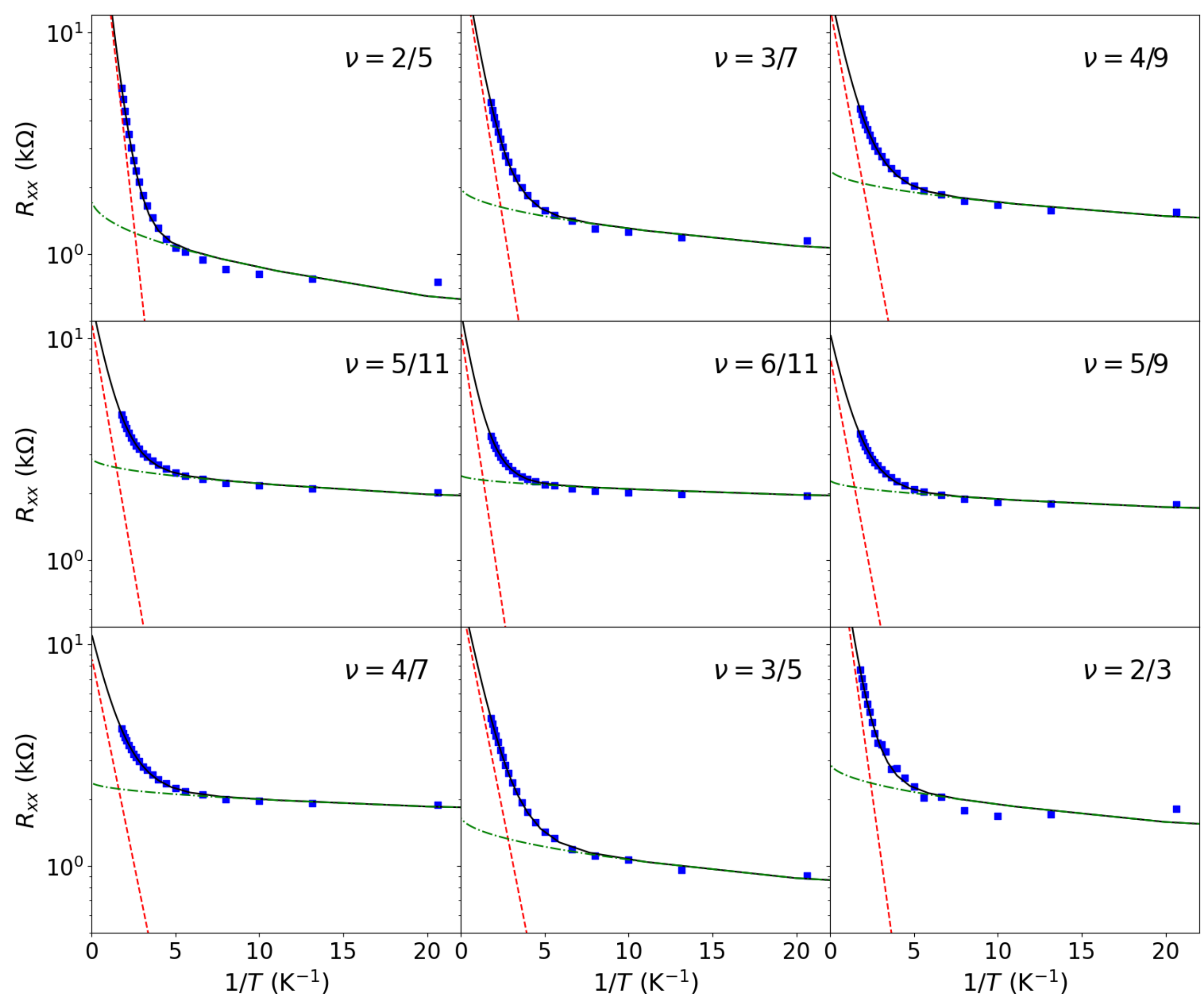

***Extended Data Figure 7. Arrhenius plots for the fractional states at zero field.*** *Arrhenius plots with thermal activation plus Efros–Shklovskii ($\beta = 1/2$) variable range hopping conduction fits for the temperature dependence of the FCI states at zero field.*

| | 2/5 | 3/7 | 4/9 | 5/11 | 1/2 | 6/11 | 5/9 | 4/7 | 3/5 | 2/3 |
|---|---|---|---|---|---|---|---|---|---|---|
| $aT^2 + b$ | 0.9785 | 0.9996 | 0.9958 | 0.9959 | 0.9948 | 0.9992 | 0.9957 | 0.9943 | 0.9963 | 0.9934 |
| $Ae^{-\Delta/2T} + Be^{-\sqrt{T_0/T}}$ | 0.9986 | 0.9989 | 0.9981 | 0.9977 | 0.9967 | 0.9973 | 0.9971 | 0.9983 | 0.9997 | 0.9947 |

***Extended Data Table 1.*** *$R^2$ values for both fitting methods for each fraction at zero field.*

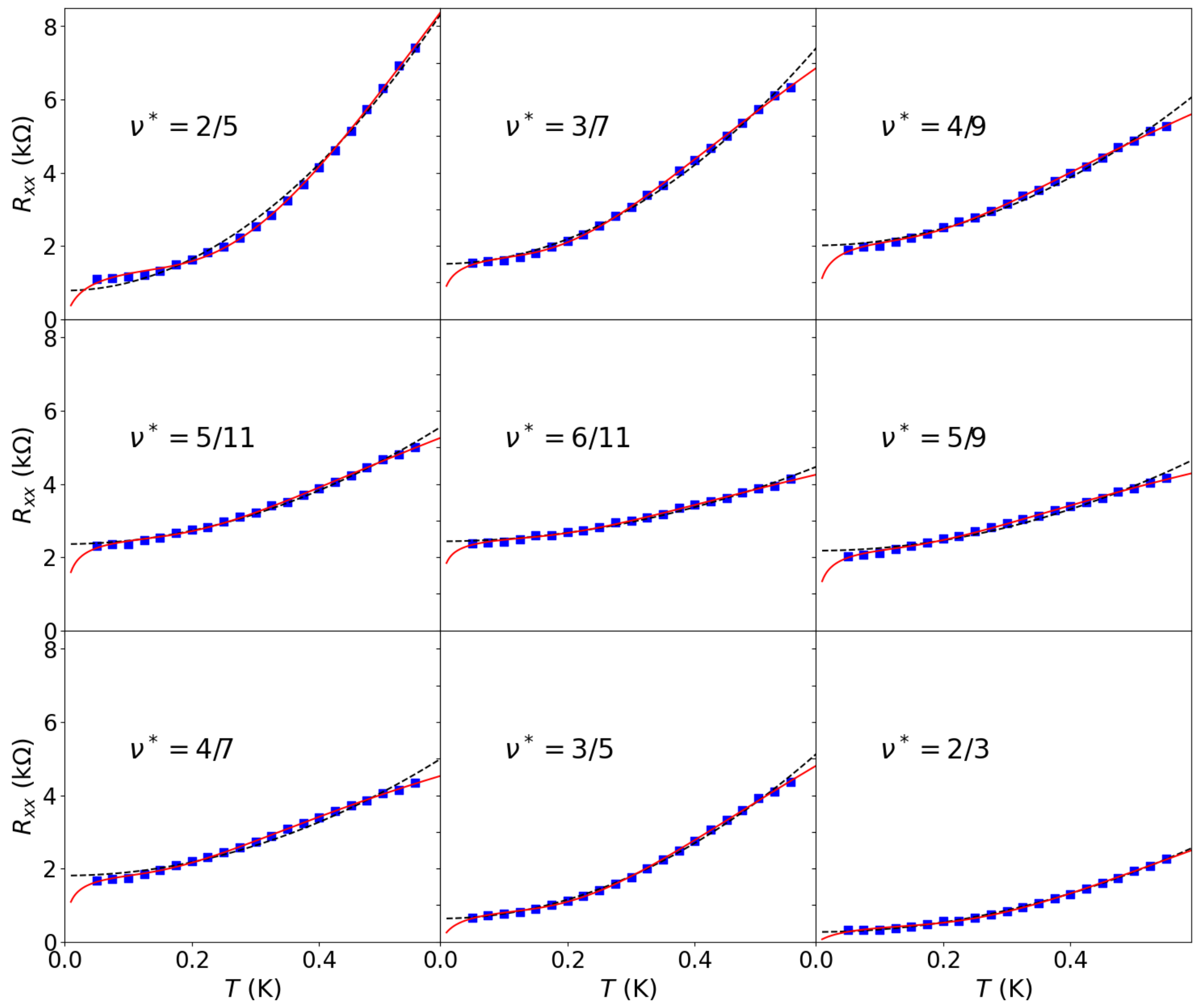

***Extended Data Figure 8. Comparison of thermally activated vs. metallic temperature dependence fits at 2 T.*** *$R_{xx}$ versus $T$ plotted on a linear scale for each FCI state at 2 T. The dashed black lines are fits to a $T^2$ law and the solid red lines are fits to a sum of thermal activation plus Efros-Shklovskii hopping conduction.*

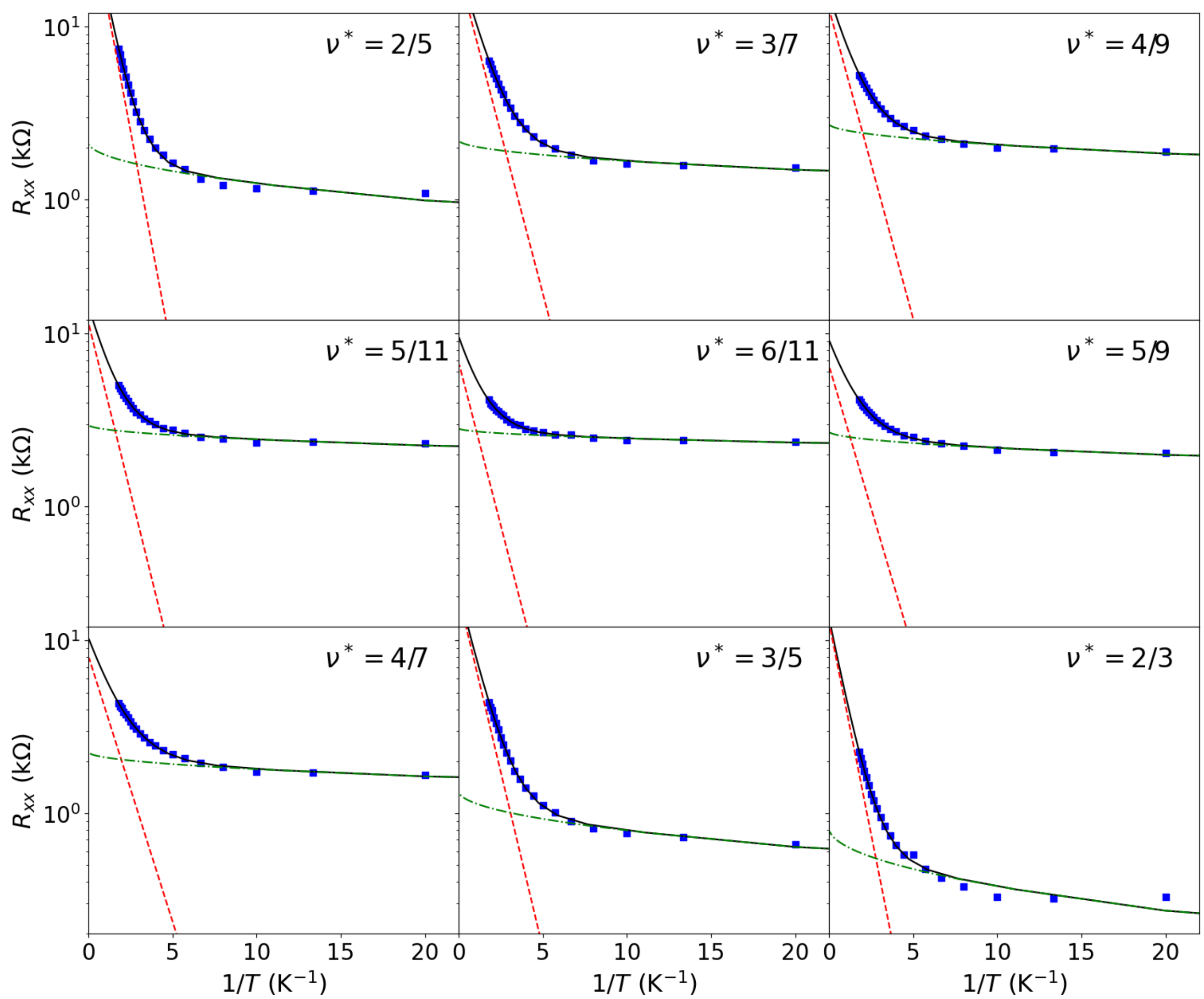

***Extended Data Figure 9. Arrhenius plots for the fractional states at 2 T.*** *Arrhenius plots with thermal activation plus Efros–Shklovskii ($\beta = 1/2$) variable range hopping conduction fits for the temperature dependence of the FCI states at 2 T.*

| | 2/5 | 3/7 | 4/9 | 5/11 | 1/2 | 6/11 | 5/9 | 4/7 | 3/5 | 2/3 |
|---|---|---|---|---|---|---|---|---|---|---|
| $aT^2 + b$ | 0.9948 | 0.9962 | 0.9911 | 0.9955 | 0.9926 | 0.9912 | 0.9800 | 0.9810 | 0.9975 | 0.9988 |
| $Ae^{-\Delta/2\mathrm{T}} + Be^{-\sqrt{T_0/T}}$ | 0.9993 | 0.9995 | 0.9989 | 0.9981 | 0.9975 | 0.9970 | 0.9979 | 0.9992 | 0.9996 | 0.9982 |

***Extended Data Table 2.*** $R^2$ values for both fitting methods for each fraction at 2 T.